\documentclass[12pt,letterpaper]{article}

\usepackage{amsmath}
\usepackage{amsfonts}
\usepackage{amssymb}
\usepackage{graphicx}
\def\be{\begin{equation}}
\def\ee{\end{equation}}

\begin{document}

\begin{flushright} {\footnotesize HUTP-A023}  \end{flushright}
\vspace{5mm}
\vspace{0.5cm}
\begin{center}

\def\thefootnote{\fnsymbol{footnote}}

{\Large \bf CMB 3-point functions generated by non-linearities at \\[.2cm] recombination} \\[1cm]
{\large Paolo Creminelli$^{\rm a}$ and Matias Zaldarriaga$^{\rm a,b}$}
\\[0.5cm]

{\small 
\textit{$^{\rm a}$ Jefferson Physical Laboratory, \\
Harvard University, Cambridge, MA 02138, USA}} 

\vspace{.2cm}

{\small 
\textit{$^{\rm b}$ Center for Astrophysics, Harvard University \\
Cambridge, MA 02138, USA
}}
\end{center}

\vspace{.8cm}

\hrule \vspace{0.3cm} 
{\small  \noindent \textbf{Abstract} \\[0.3cm]
\noindent
We study the 3-point functions generated at recombination in the squeezed triangle limit, when one mode has  a wavelength  much larger than the other two and 
is outside the horizon. The presence of the long wavelength mode cannot change the physics inside the horizon but modifies how a late time observer sees the anisotropies. 
The effect of the long wavelength mode can be divided into a redefinition of time and spatial scales, a Shapiro time delay and gravitational lensing. The separation 
is gauge dependent but helps develop intuition.  We show that the resulting 3-point function corresponds to an $f_{\rm NL} < 1$ and that its shape is different 
from that  created by the $f_{\rm NL}$ (or local) model. 
   
\vspace{0.5cm}  \hrule

\def\thefootnote{\arabic{footnote}}
\setcounter{footnote}{0}

\section{Introduction}

Conventional slow-roll inflation predicts that the primordial perturbations are Gaussian to an extreme level of accuracy, with deviations of order 
$10^{-6}$ \cite{Maldacena:2002vr,Acquaviva:2002ud}. Any departure from the simplest picture, like the addition of light scalars during inflation or models with modified inflaton dynamics, leads to an enhancement of the non-Gaussianity and many of these models predict a 3-point function which should be detected by the forthcoming experiments \cite{Lyth:2002my,Zaldarriaga:2003my,Creminelli:2003iq,Arkani-Hamed:2003uz,Alishahiha:2004eh}. 

The relation between the primordial fluctuations and what we observe today is in general non-linear. Although the linear approximation is quite good for studying the 
CMB anisotropies on large enough scales, small non-linearities will generate higher order moments even starting from a completely Gaussian primordial spectrum. 
The leading sources of non-Gaussianity are related to secondary anisotropies, anisotropies imprinted after recombination $(z\sim 1100)$. These anisotropies can be 
divided into scattering secondaries, when the CMB photons scatter with electrons along the line of sight, and gravitational secondaries when effects are mediated by 
gravity (for a review of all the different secondary effects see \cite{Hu:2001bc}).  Examples of scattering secondaries are the thermal Sunyaev-Zeldovich effect, where 
hot electrons in clusters and other collapsed objects transfer energy to the CMB photons, the kinetic Sunyaev-Zeldovich effect produced by the bulk motion of the 
electrons in those clusters, the Ostriker-Vishniac effect, produced by bulk motions modulated by linear density perturbations, and effects due to the patchy nature of 
the reionization process. The various correlations between these different effects lead to non-zero 3-point functions and other higher order effects (see for example 
\cite{Cooray:1999kg}).  As the growth of structure proceeds, density inhomogeneities, bulk and thermal motions  grow and become quite large on small length scales. 
As a result the scattering secondaries are produced mostly at late times and are most significant on small angular scales.  

Gravitational secondaries arise from two separate effects, the change in energy of photons when the gravitational potential is time-dependent 
and gravitational lensing.  Gravitational secondaries are different from scattering secondaries in that even as structure formation evolves the depth of 
gravitational potential wells does not grow significantly, always being of order $10^{-5}$ (\footnote{The potential wells are smaller on scales that were inside 
the horizon during the radiation era as fluctuations are not able to grow during this period (see for example \cite{peebles})}). At late times, when the 
Universe becomes dominated by the cosmological constant the gravitational potential on linear scales starts to decay. This leads to the largest
of the gravitational secondaries, the integrated Sachs-Wolfe effect (ISW).   The ISW effect is linear in the gravitational potential and affects mainly 
large angular scales. Other secondaries that result from a time dependent potential are the Rees-Sciama effect, produced by the time 
evolution of the potential on non-linear scales, and the moving cluster effect, where time-dependence is produced by motions. Both these effects are subdominant.  

The fact that the potential never grows appreciably means that most second order effects created by gravitational secondaries are subdominant to those created by 
scattering ones. The exception are effects related to gravitational lensing. The reason is that the total defection angle of a photon on its way from the last scattering surface can be more than an order of magnitude larger than $\phi$. This is because  for a mode of wavenumber $k$, there are $k/H$ independent regions along  the line of sight. Gravitational lensing conserves surface brightness so it does not create anisotropies, it only modifies existing ones. Lensing  is 
thus a second order effect but the resulting anisotropies are parametrically larger than $\phi^2$. 

The most important  3-point function induced by secondaries on large angular scales is the one produced by the correlation between the gravitational lensing effect 
and the ISW effect. The matter distribution along the line of sight both lenses the primary anisotropies and creates additional ones through the ISW effect. This 
results in a non-zero 3-point function first calculated in \cite{Seljak:1998nu,Goldberg:xm}. Due to the enhanced nature of the gravitational lensing effect, this 
3-point function is rather large and is expected to be detected by the next generation of full sky CMB experiments such as Planck.  

A second source of non-Gaussianity are non-linearities operating at recombination.  Not much is known about the explicit form of their resulting 3-point function. 
The naive estimate is that higher order corrections are suppressed by powers of the Newtonian potential $\phi$, so that non-Gaussianities should be of order 
$3 \cdot 10^{-5}$. It is quite important to refine this estimate. First of all this estimated level of non-Gaussianity is roughly of the same order as what the 
forthcoming Planck mission can detect; this implies that to fully exploit Planck data we must be able to control the 3-point function generated at recombination. 
Moreover it is not {\em a priori} clear that the naive estimate of a $10^{-5}$ non-Gaussianity is a good one; physics at recombination is quite complicated and rich 
(think about the structure of the acoustic peaks in the various spectra) and there could be a sizeable suppression or enhancement.

The calculation of the full 3-point function of the primary anisotropies is very complicated. One should take into account all the non-linearities in the evolution of
the baryon-photon fluid at recombination and  non-linearities coming from general relativity.  The purpose of this paper is more modest. We are going to derive 
the result in a particular geometrical limit for the 3-point function. In Fourier space we take one of the modes to be of very long wavelength, in such a way 
that it is outside the horizon at recombination. Moreover we take the other two modes to be of much smaller wavelength, so that the triangle in Fourier space 
is very squeezed, with one side much smaller than the others.

There is a simple physical reason why the calculation is much simpler in this limit. At recombination the long wavelength mode is outside the horizon 
so that it cannot change how recombination happens; it is completely irrelevant before it reenters in the horizon. After reentering the long
wavelength mode is physical, so that the 2-point function of the small wavelength modes will be different moving along the long mode and this 
will generate a 3-point function. The background mode cannot change the features of the 2-point function which encode physics at recombination 
(for example the ratio between the first and second peek) but it can only change the angle under which a given physical scale is observed. Moving 
along the long wave the 2-point function will be stretched or shrunk (generically in an anisotropic way), but will preserve its features. 

In the following section we will study the effect of a long wavelength mode on the 2-point function. As the 2-point function is already at first order in the
perturbation we have to consider the effect of the background wave at first order only. There are different effects but, even though it is simpler 
and more intuitive to study them separately, the separation among them is gauge dependent and therefore non-physical. Once we know at leading order in 
the perturbation the effect of the long-wavelength mode, we can calculate the 3-point function for the temperature fluctuations (section \ref{sec:3p}). The 
discussion is easily extended to polarization spectra (section \ref{sec:pol}). The significance of our results and the comparison with 
related works are discussed in the conclusions (section \ref{sec:conc}).

\section{\label{sec:back}The effect of the background wave}
A useful way to write the metric at non-linear order in the presence of scalar perturbations outside the horizon is 
\begin{equation}
\label{eq:exp}
ds^2 = a^2(\eta) \left[-d\eta^2 + e^{2\zeta(x^i)} dx_i dx_i\right] \;.
\end{equation}
The variable $\zeta$ is constant outside the horizon \cite{Salopek:1990re,Maldacena:2002vr} because, in the limit in which we can neglect gradients, the effect 
of the exponential is just an irrelevant rescaling of the spatial coordinates. It has been shown \cite{Maldacena:2002vr} that the 3-point function of {\em this}
variable in standard slow-roll inflation is very small, being suppressed by a linear combination of the slow-roll parameters:
$\left\langle\zeta\zeta\zeta\right\rangle \sim {\cal{O}}(\epsilon,\eta) \left\langle\zeta\zeta\right\rangle^2$.
In the following we are going to neglect this original non-linearity as it is much smaller than the effects we are going to calculate,
which come from the non-linear relation between the temperature fluctuation we observe and $\zeta$.
One could prefer to use another non-linear expression for $\zeta$, for example the variable $\zeta_{\rm new}(x) = \zeta(x) + \zeta(x)^2$. 
This new variable would obviously still be conserved outside the horizon. But now the 3-point function of 
$\zeta_{\rm new}$ is not suppressed by the slow-roll parameters and it would have to be taken into account because of the same order of the effects 
we are going to calculate. Obviously the final physical result, {\em i.e.} the 3-point function of the temperature fluctuations, does not depend
on our choice of variables, but the separation between what is ``created'' during inflation and what derives from the non-linear relation between 
the metric and what we observe does.

There is another reason why the exponential form (\ref{eq:exp}) is useful. Suppose we want to study, as discussed in the Introduction, the effect
of a long background mode which is outside the horizon on smaller scale fluctuations: $\zeta = \zeta_{\rm long} + \zeta_{\rm short}$. Only
with the exponential form the background mode effect multiplies the short modes and it is thus equivalent to an unobservable coordinate rescaling, before the long mode
finally reenters in the horizon. This is necessary for our intuitive arguments in the Introduction. 
Taken this useful choice of variables, now we are going to discuss the effect of the background mode. We do it at linear order because it will affect
the 2-point function of the short modes which is already at first order in the perturbation. 

When the long wavelength mode is still outside the horizon, the metric can be written in terms of the Bardeen variable $\zeta$ as
\begin{equation}
\label{eq:comoving}
ds^2 = a^2(\eta) \left[-d\eta^2 + (1+2\zeta(x^i))dx_i dx_i\right] \;,
\end{equation}
where $\zeta$ is time independent and slowly varying as a function of $x^i$ compared to the Hubble scale. This metric is exact up to corrections
subleading in the $k/H$ expansion. As the mode is outside the horizon, recombination
happens {\em exactly} in the same way in different points along the $\zeta$ wave. Moreover, as these coordinates are comoving, recombination
happens at the same conformal time $\eta=\eta_r$ everywhere\footnote{We are not assuming that recombination is instantaneous. Whatever is the time dependence
of the process it will be the same everywhere, independently of the position along the long-wave mode.}. The only effect of the background wave is to rescale 
the spatial coordinates, so that at different points along the $\zeta$ wave the same physical scale ({\em e.g.} the Hubble horizon) will be expressed 
differently in terms of $x^i$:
\begin{equation}
\label{eq:physical}
a(\eta_r) (1+\zeta(x^i)) \Delta x = {\rm physical \;scale} = {\rm const.} 
\end{equation}

To follow the long wave mode inside the horizon and to compare with the existing literature we go to conformal Newtonian gauge 
when the mode is still outside the horizon. The metric is given by
\begin{equation}
\label{eq:Newton}
ds^2 = a^2(\tau) \left[-(1+2 \phi(x^i))d\tau^2 + (1-2\phi(x^i))dx_i dx_i\right] \;,
\end{equation}
with the well-known relation $\phi = -\frac35 \zeta$ during matter dominance. In the long wavelength limit there is no redefinition of the 
spatial coordinates going from eq.~(\ref{eq:comoving}) to (\ref{eq:Newton}). Therefore the $x$ expression of a fixed physical distance depends on 
the position on the background wave as in eq.~(\ref{eq:physical}).  

In Newtonian gauge there is an additional effect to take into account. Recombination does not occur at the same conformal time $\tau$ everywhere.
As in matter dominance $a(\tau) \propto \tau^2$ the relation between the conformal times in the two gauges is given by
\begin{equation}
\label{eq:timerep}
\tau = \big(1-\frac13 \phi(x)\big) \eta \;.
\end{equation}
As recombination happens at a given $\eta = \eta_r$ everywhere, the expression in Newtonian gauge will be given by 
\begin{equation}
\label{eq:timerec}
\tau_r(x) = \big(1-\frac13 \phi(x)\big) \eta_r \;.
\end{equation}

We are now ready to calculate the effect of the long wavelength mode on the 2-point function. Although physics is locally
unchanged at recombination by the presence of the long wavelength mode, this background wave will change the angular scale at 
which the 2-point function is observed. The final angle of observation is changed by the $\phi$ wave both because the  
expression in the $(\tau, x)$ coordinates depends on $\phi$ and because of the geodesic propagation in the perturbed metric from the
last scattering surface to us.
The separation between these two effects is actually gauge dependent: only the total deviation is physical and therefore gauge independent. 
We will discuss the issue of gauge dependence in Appendix A.
As we are interested in the angular redefinition at linear order in $\phi$ we can consider each effect separately. We start from the 
$(\tau, x)$ redefinition.

{\bf Time redefinition.} Both time and space redefinition change the angular scale at which a given mode is observed. The angular scale is given by 
\begin{equation}
\label{eq:angle}
\theta = \frac{\Delta x}{\tau_0-\tau_r} \;,
\end{equation}
where $\Delta x$ is a given separation on the last scattering surface and $\tau_0 - \tau_r$ is the distance to the last scattering surface.
The distance to the last scattering surface is dominated by the present conformal time $\tau_0$, so that
the fluctuations of order $\phi$ in the conformal time of recombination of eq.~(\ref{eq:timerec}) change the angular scale only by
\begin{equation}
\label{eq:dteta1}
\frac{\delta\theta}{\theta} \sim \frac{\tau_r}{\tau_0} \phi \sim \frac1{(1+z_r)^{1/2}} \phi \;, 
\end{equation} 
where $z_r \sim 1100$ is the red-shift at recombination. This effect is therefore much smaller than the $\Delta x$ fluctuations 
of eq.~(\ref{eq:physical}) which give a relative angular variation of order $\phi$. Fluctuations in $\tau_r$
can therefore be neglected.

{\bf Space redefinition.} The long wavelength mode, redefining distances as in eq.~(\ref{eq:physical}), gives an isotropic ({\em i.e.} independent of 
the orientation with respect to the background mode) redefinition of the angle of observation
\begin{equation}
\label{eq:anglered}
\frac{\delta \theta}{\theta} = \frac53 \phi \;.
\end{equation}
Going to Fourier space, this implies a variation of the $C_l$'s of the form
\begin{equation}
\label{eq:Clred}
\delta (l^2 C_l) = - \frac53 \phi \cdot \frac\partial{\partial \log l} (l^2 C_l) \;.
\end{equation}
For a scale invariant spectrum $C_l \propto l^{-2}$ there is obviously no effect.

Let us now consider the effect of the propagation on the perturbed background. There are two effects. The first one is the variation of the position
of the last scattering surface due to the potential along the line of sight (Shapiro time delay), the second is the variation of the angle due
to the non-homogeneity of the background field (lensing).

{\bf Shapiro time delay.} To study the effect of the Shapiro time delay we take the equation for the geodesic in the metric (\ref{eq:Newton}) at 
zeroth order in the spatial derivatives of $\phi$, {\em i.e.} neglecting lensing effects
\begin{equation}
\label{eq:geodesics}
\tau - \tau_r(x) = \int (1-2\phi(x')) dx' \;.
\end{equation}    
This implies that the radial position of the last scattering surface is not only influenced by the different time of emission $\tau_r(x)$ of 
eq.~(\ref{eq:timerec}), but also by the Newtonian potential $\phi(x)$ {\em along the line of sight}. This effect has been considered in \cite{Hu:yq}. 
However, the effect is suppressed with respect to the fluctuations in $\Delta x$ discussed above. The reason is that the 
integral of $\phi$ in the right hand side of eq.~(\ref{eq:geodesics}) tends to average to zero unless the mode wavevector is perpendicular to the line of sight. 
This implies that this effect will be comparable to eq.~(\ref{eq:physical}) only for the lowest multiples.

{\bf Lensing.} A spatially inhomogeneous Newtonian potential $\phi$ will deflect the geodesics and give rise to lensing. 
In the Born approximation the shear matrix $\Phi_{ij}$ can be written as an integral over the unperturbed line of sight \cite{Seljak:1995ve}
\begin{equation}
\label{eq:shear}
\Phi_{ij} \equiv \frac{\partial \delta \theta_i}{\partial \theta_j} = 2 \int_0^{D_{\rm LSS}} \!\!\!\!d D \;\frac{D(D_{\rm LSS}-D)}{D_{\rm LSS}} 
\nabla_i \nabla_j \phi(D) \;,
\end{equation}
where $D_{\rm LSS}$ is the comoving distance to the last scattering surface, which we take infinitely thin. Although we have a certain cancellation
along the line of sight as for the Shapiro time delay, the additional gradients acting on the Newtonian potential enhance the lensing effect. In the next section 
we will show that lensing gives a contribution to the 3-point function comparable to the space redefinition effect.  
The part of the shear matrix proportional to the identity matrix gives rise to an isotropic rescaling of the angles, similarly to what
happens in eq.~(\ref{eq:anglered}). In this case the relative variation of the angle from the last scattering surface to us is given by the 
convergence $\kappa$
\begin{equation}
\label{eq:converg}
\frac{\delta \theta}{\theta} = \kappa \equiv \int_0^{D_{\rm LSS}} \!\!\!\!d D \;\frac{D(D_{\rm LSS}-D)}{D_{\rm LSS}} 
\nabla^2_\perp \phi(D) \;,
\end{equation}
where $\nabla^2_\perp$ is the Laplacian restricted to the directions perpendicular to the line of sight. Contributions to the shear matrix not proportional
to the identity matrix will give an anisotropic redefinition of the angle of observation, {\em i.e.} dependent on the orientation with respect to the
background wave. 

Before going on an important remark. We are going to consider the 3-point function induced by lensing, correlating the lensing effect of the background wave
with the temperature fluctuations it produces close to the last scattering surface. As we mentioned in the introduction, in a Universe with a cosmological constant 
the Newtonian potential is not constant and an additional contribution to the temperature fluctuation comes from the Integrated Sachs Wolfe effect. It is well known 
\cite{Seljak:1998nu,Goldberg:xm} that the correlation of this effect along the line of sight with lensing gives quite a  big 3-point function, which is sensitive 
to the value of the cosmological constant. This effect is physically distinct from what we are studying as it occurs along the line of sight and it could, 
at least in principle, be separated through an independent reconstruction of the projected potential for example by studying weak lensing of galaxies.

\section{\label{sec:3p}The 3-point function}
In this section we explicitly present the 3-point function for the temperature fluctuations coming from the effects discussed above, in the flat
sky limit. This limit is not a good approximation if the background mode has a wavelength comparable to our present horizon, 
{\em i.e.} for its lowest multiples. Nevertheless the expressions are much more transparent in this limit so we leave for appendix B the full
expressions for the spherical geometry.

We define the spectrum of the 2-point function in the flat sky limit as
\begin{equation}
\label{eq:spectrum}
\Big\langle \frac{\delta T(\vec l_2)}{T} \frac{\delta T(\vec l_3)}{T} \Big\rangle = (2 \pi)^2 \delta^{(2)} (\vec l_2 + \vec l_3) C_{l_2} \;.
\end{equation}
In a similar way we can factor out from the 3-point function the delta of momentum conservation defining
\begin{equation}
\label{eq:defF}
\Big\langle \frac{\delta T(\vec l_1)}{T} \frac{\delta T(\vec l_2)}{T} \frac{\delta T(\vec l_3)}{T} \Big\rangle 
= (2 \pi)^2 \delta^{(2)}(\vec l_1 + \vec l_2 + \vec l_3) F(\vec l_1 ; \vec l_2 ; \vec l_3 ) \;.
\end{equation}

As we discussed in the previous section, the two leading effects of the background wave on the 2-point function are space redefinition and lensing.
The 3-point function from the space redefinition effect is obtained correlating the temperature fluctuation given by the long wavelength mode with its effect 
on the 2-point function  in eq.~(\ref{eq:Clred}). We get
\begin{equation}
\label{eq:3-point1}
F(\vec l_1 ; \vec l_2 ; \vec l_3)
= \Big\langle \frac{\delta T(\vec l_1)}{T} \;\phi(\vec l_1)^* \Big\rangle^\prime
 \frac1{l_2^2} \left(-\frac53 \frac{\partial}{\partial \log l} (l_2^2 C_{l_2})\right) \;,
\end{equation}
where with the prime we indicate the corresponding quantity without the $(2\pi)^2\delta^{(2)}$ of momentum conservation. Let us stress
again what is the range of validity of this expression. The ``background'' mode $\vec l_1$ must be outside the horizon at recombination and we must be in 
squeezed, ``Maldacena'' \cite{Maldacena:2002vr} limit
\begin{equation}
\label{eq:validity}
l_1 \ll 200 \;,\qquad\qquad  l_2 \simeq l_3 \gg l_1 \;.
\end{equation} 
An expression similar to eq.~(\ref{eq:3-point1}) holds for the isotropic lensing contribution
\begin{equation}
\label{eq:3-point2}
F(\vec l_1 ; \vec l_2 ; \vec l_3 )
= \Big\langle \frac{\delta T(\vec l_1)}{T} \;\kappa(\vec l_1)^* \Big\rangle^\prime
\frac1{l_2^2} \left(-\frac{\partial}{\partial \log l} (l_2^2 C_{l_2})\right) \;.
\end{equation}
Note that these two contributions are quite similar and we expect a partial cancellation between the two.  The reason is that regions close
to the last scattering surface with positive convergence $\kappa$ (\footnote{In the integral of eq.~(\ref{eq:converg})
only the contribution close to the last scattering surface has a substantial correlation with the temperature $\delta T(\vec l_1)$ of the long
wavelength mode.}) are regions of overdensity, as the Laplacian of $\phi$ is proportional
to the mass overdensity. Regions of overdensity are regions of negative Newtonian potential $\phi$ and this implies a partial cancellation between 
eq.s (\ref{eq:3-point1}) and (\ref{eq:3-point2}).

Let us now consider the anisotropic lensing contribution. The background plane wave depends only on one direction in the sky, so that in a suitable basis 
the shear matrix will be of the form\footnote{This is not true for a generic background, like for example a spherical harmonic.}
\begin{equation}
\label{eq:flatshear}
\Phi_{ij} = \left(\begin{array}{cc} 2\kappa & 0 \\ 0 & 0 \end{array} \right) \;.
\end{equation}
This implies that the anisotropic shear is 
\begin{equation}
\label{eq:anyshear}
\Phi_{ij} ^{\rm anis}= \left(\begin{array}{cc} \kappa & 0 \\ 0 & -\kappa \end{array} \right) \;.
\end{equation}
The angle of observation of a given short mode at recombination is redefined in a way that is proportional to $\cos(2\varphi_{12})$, where $\varphi_{12}$ 
is the angle between the short mode and long lensing wave. This gives us a 3-point function of the form 
\begin{equation}
\label{eq:3-point3}
F(\vec l_1 ; \vec l_2 ; \vec l_3 )
=  \Big\langle \frac{\delta T(\vec l_1)}{T} \;\kappa(\vec l_1)^* \Big\rangle^\prime
\cos(2 \varphi_{12})\frac1{l_2^2} \left(-\frac{\partial}{\partial \log l} (l_2^2 C_{l_2})\right) \;.
\end{equation}
 
For any 3-point function we can calculate the signal to noise ratio $S/N$ for an ideal cosmic variance limited experiment. It is given in the flat sky 
approximation by \cite{Hu:ee}
\begin{equation}
\label{eq:SN}
(S/N)^2 = \frac1\pi \int \frac{d^2 l_1 d^2 l_2}{(2 \pi)^2} \frac{F(\vec l_1 ; \vec l_2 ; \vec l_3)^2}{6 C_{l_1}C_{l_2}C_{l_3}} \;.
\end{equation}
For the space redefinition contribution of eq.~(\ref{eq:3-point1}) the quantity 
$\Big\langle \frac{\delta T(\vec l_1)}{T} \;\phi(\vec l_1)^* \Big\rangle^\prime$ can be expressed in terms of the power spectrum $P_{\cal{R}}$
using the Sachs-Wolfe formula $\delta T/T = \phi/3$, which is valid for modes which are outside the horizon at recombination\footnote{We are neglecting
the contribution of the Integrated Sachs-Wolfe effect. This effect is created by the variation with time of the potential close to us, at low red-shift.
It is therefore correlated to the potential on the last scattering surface only for the very low multiples.}
\begin{equation}
\label{eq:phiT}
\Big\langle \frac{\delta T(\vec l_1)}{T} \;\phi(\vec l_1)^* \Big\rangle^\prime = 
\frac3{25} \langle\zeta(\vec l_1) \zeta(\vec l_1)^* \rangle^\prime =\frac3{25} \frac{2\pi}{l_1^2} P_{\cal R} \;.
\end{equation}
The best constraint on the power spectrum $P_{\cal R}$ now comes from the WMAP collaboration $P_{\cal R}^{1/2} \simeq 4.3 \times 10^{-5}$
\cite{Peiris:2003ff}. The logarithmic derivative of the short wavelength spectrum can be calculated using publicly available codes like CMBFAST \cite{Seljak:1996is}.
In figure \ref{fig:SN1} we show the signal to noise ratio as a function of the maximum $l$ used for the short modes in an ideal experiment. We restricted the
sum to configurations satisfying (\ref{eq:validity}) only. 

Let us now concentrate on eq.~(\ref{eq:3-point2}). In the expression for the correlation between the temperature fluctuation and the convergence of the long 
wavelength mode we have been sloppy. As the convergence is an integral over the line of sight there is no one to one correspondence between a mode we see in the sky and
the momentum perpendicular to the line of sight. This implies that a given mode $\vec l_1$ of $\delta T/T$ is not exclusively correlated to a mode $-\vec l_1$ of the 
convergence as implicit in eq.~(\ref{eq:3-point2}). Nevertheless we have checked that this effect is quite small, except for the lowest multiples, as the temperature 
fluctuation is correlated almost exclusively with the contribution to the convergence close to the last scattering surface. In this limit  
\begin{equation}
\label{eq:kappaphi}
\Big\langle \frac{\delta T(\vec l)}{T} \;\kappa(\vec l)^* \Big\rangle^\prime = -\frac3{25} P_{\cal{R}} \int \frac{d k_z}{2\pi} 
\frac{2 \pi^2}{(k_z^2 + l^2/D_{\rm LSS}^2)^{3/2}} \int_0^{D_{\rm LSS}} \frac{d D}{D_{\rm LSS}} \frac{D}{D_{\rm LSS}}(1-\frac{D}{D_{\rm LSS}}) \frac{l^2}{D_{\rm LSS}^2}
e^{i k_z (D-D_{\rm LSS})} \;,
\end{equation}
where we have approximated $D = D_{\rm LSS}$ is some place.
From this we get 
\begin{equation}
\label{eq:kappaphi2}
\Big\langle \frac{\delta T(\vec l)}{T} \;\kappa(\vec l)^* \Big\rangle^\prime = \frac3{25} (- 2 \pi l^2) P_{\cal{R}} \int_0^\infty dy 
\frac1{(y^2+l^2)^{3/2}} \int_0^1 dx \; x (1-x) 
\cos(y (1-x)) \;,
\end{equation}
where the integral is easy to evaluate numerically. What is the dependence on $l$ of this expression? There is an $l^2$ enhancement with respect to the
correlator (\ref{eq:phiT}) coming from the Laplacian in the lensing expression. This seems to suggest, neglecting the lowest multiples, that lensing 
should be dominant. Actually the $l^2$ enhancement is cancelled by two other effects. The first one is the cancellation along the line of sight we already
discussed for the Shapiro time delay; if the wavevector has a component along the line of sight the background wave will lens the 2-point function 
in opposite ways along the path with a resulting partial cancellation. The second effect comes from the window function $D (D_{\rm LSS}-D)/D_{\rm LSS}$ 
in expression (\ref{eq:shear}). This tells us that lensing close to the last scattering surface is very ineffective, as it is easy to understand geometrically.
The correlation between lensing and temperature fluctuation is therefore suppressed. It can be explicitly checked in the integral (\ref{eq:kappaphi2}) that 
both these effects give an $l^{-1}$ suppression, so that the correlator (\ref{eq:kappaphi2}) finally behaves like the one in eq.~(\ref{eq:phiT}), it
goes as $l^{-2}$. Space redefinition and lensing give comparable contributions.

Expression (\ref{eq:kappaphi2}) allows us to estimate $S/N$ for the 3-point function of eq.~(\ref{eq:3-point2}). 
In figure \ref{fig:SN1} we show the result as a function of the maximum $l$ for an ideal experiment. We also show $S/N$ for the sum of the two effects discussed
so far to show that there is a partial cancellation between the two.

\begin{figure}[!!ht]             
\begin{center}
\includegraphics[width=12cm]{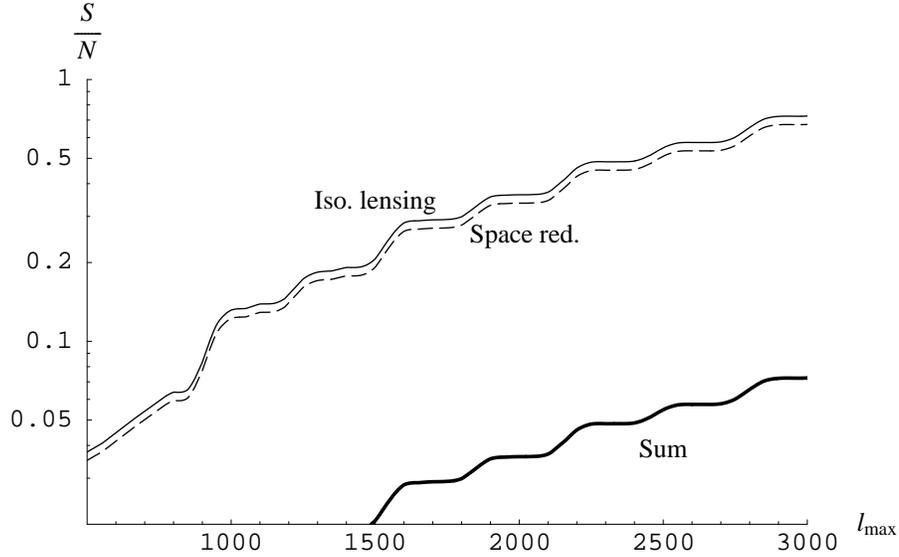}
\caption{\label{fig:SN1} \small Signal to noise as a function of the maximum $l$ for an ideal, cosmic variance limited experiment. Dashed line: space redefinition effect.
Continuous thin line: isotropic lensing effect. Thick line: total effect. We see that the total signal to noise is smaller than the individual 
contributions as the two effects partially cancel. The parameters of the cosmological model we used were: $\Omega_b=0.045$, $\Omega_c=0.255$, $\Omega_\Lambda=0.7$,  $h=0.7$ and $\sigma_8=0.85$.}
\end{center}
\end{figure}

For the anisotropic lensing, it is easy to relate its signal to noise ratio to the one given by the isotropic lensing effect. Averaging expression
(\ref{eq:3-point3}) over the orientation between the background mode and the 2-point function we get a signal to noise ratio smaller by a factor
of $\sqrt{2}$ with respect to the isotropic lensing effect. We plot the result in fig.~\ref{fig:SN2}. As a consequence of the partial cancellation
between the isotropic effects, the anisotropic lensing gives the biggest signal to noise.

\begin{figure}[!!ht]             
\begin{center}
\includegraphics[width=12cm]{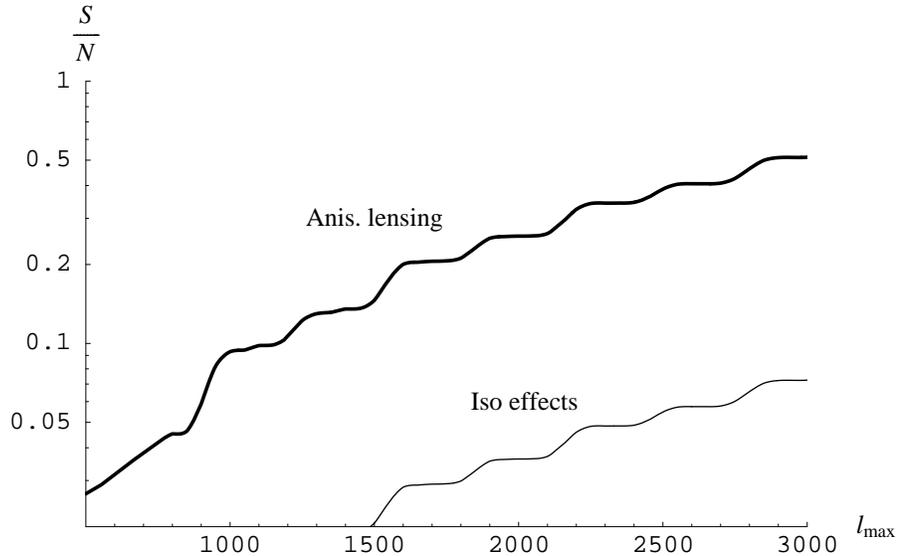}
\caption{\label{fig:SN2} \small Signal to noise for the anisotropic lensing contribution (thick line) compared to the sum of the isotropic contributions (thin line)
as a function of the maximum $l$ for an ideal, cosmic variance limited experiment. The anisotropic effect has a bigger signal to noise.
Cosmological parameters are the same as in the previous figures.}
\end{center}
\end{figure}

\section{\label{sec:pol}Polarization 3-point functions.}
We can easily extend the results of the previous sections to the 3-point functions involving polarization modes. Note first of all that 
the $E$ polarization decays quite fast for modes outside the horizon at recombination, while $B$-modes are zero in absence of a tensor contribution. 
This implies that, in the limit we are considering 
({\em i.e.} one long-wavelength mode outside the horizon at recombination and two much shorter modes), the long mode must be a temperature mode. 
Therefore we have only to consider four 3-point functions involving polarization modes: $TTE$, $TEE$, $TTB$, $TEB$, where the first mode is the 
long-wavelength mode. 

The discussion for the $TTE$ and $TEE$ modes follows closely what we said about the 3-point function of the temperature map. The effect of the background mode
is always to redefine the angle at which a certain physical scale is observed. Given the partial cancellation discussed above of the two
isotropic effects (space redefinition and isotropic lensing) the leading effect is the anisotropic lensing. The explicit form for the $TTE$ and $TEE$
correlators is the same as in eq.~(\ref{eq:3-point3}), with the only difference that the small scale spectrum $C_{l_2}$ is respectively the 
$TE$ and the $EE$ spectrum:
\begin{equation}
\label{eq:3-pointE}
F_{TXE}(\vec l_1 ; \vec l_2 ; \vec l_3 )
=  \Big\langle \frac{\delta T(\vec l_1)}{T} \;\kappa(\vec l_1)^* \Big\rangle^\prime
\cos(2 \varphi_{12})\frac1{l_2^2} \left(-\frac{\partial}{\partial \log l} (l_2^2 C_{l_2}^{XE})\right) \;,
\end{equation}
where $X$ stands for either $T$ or $E$.

It is straightforward to calculate the signal to noise of these correlators from eq.~(\ref{eq:SN}). 

\begin{equation}
\label{eq:SNE}
(S/N)_{TTE}^2 = \frac1\pi \int \frac{d^2 l_1 d^2 l_2}{(2 \pi)^2} \frac{F_{TTE}(\vec l_1 ; \vec l_2 ; \vec l_3)^2}{2 C^{TT}_{l_1}C^{TT}_{l_2}C^{EE}_{l_3}} \qquad 
(S/N)_{TEE}^2 = \frac1\pi \int \frac{d^2 l_1 d^2 l_2}{(2 \pi)^2} \frac{F_{TEE}(\vec l_1 ; \vec l_2 ; \vec l_3)^2}{2 C^{TT}_{l_1}C^{EE}_{l_2}C^{EE}_{l_3}} \;.
\end{equation}
In calculating the variances at denominator in the first expression we neglect contributions involving $(C^{TE}_l)^2$ because it is much smaller than 
$C^{TT}_l\cdot C^{EE}_l$, {\em i.e.} the correlation between $E$ and $T$ is not perfect and it becomes quite small at large $l$. 

The results are given in figure \ref{fig:SN3} as a function of the maximum $l$ for an ideal cosmic variance limited
experiment. 

\begin{figure}[!!ht]             
\begin{center}
\includegraphics[width=12cm]{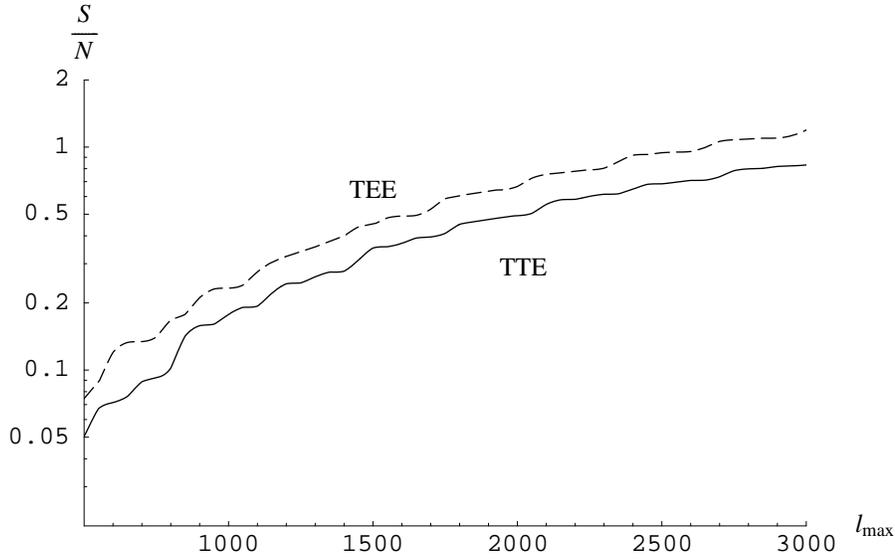}
\caption{\label{fig:SN3} \small Signal to noise for the TTE (continuous line) and TEE (dashed line) correlators as a function of the maximum $l$ for an ideal, 
cosmic variance limited experiment. Cosmological parameters are the same as in the previous figures.}
\end{center}
\end{figure}

We can finally discuss correlators involving $B$ modes. Even if we do not have any original $B$ mode from a tensor component, the anisotropic lensing
generates a $B$ component from a pure $E$ polarization. Lensing does not change the state of polarization of a given photon, but changes the angle at which 
we observe a given physical scale. The magnitude and orientation of the wavevector of a given mode is therefore changed. A pure $E$ mode is converted into a 
mixing of the two polarization states. Taking an $E$ mode in the flat sky limit it is straightforward to see that the amount of $B$ polarization
generated by anisotropic lensing is given by
\begin{equation}
\label{eq:Bmode}
B = - E \kappa \sin (2\varphi_{12}) \;,
\end{equation}
where $\kappa$ is the convergence and $\varphi_{12}$ is the angle between the lensing long-wavelength mode and the short $E$ mode.

From this we get the expression for the $TTB$ and $TEB$ correlators
\begin{eqnarray}
\label{eq:3-pointB}
F_{TTB}(\vec l_1 ; \vec l_2 ; \vec l_3 )
& = & -\Big\langle \frac{\delta T(\vec l_1)}{T} \;\kappa(\vec l_1)^* \Big\rangle^\prime
\sin(2 \varphi_{12}) C_{l_2}^{TE} \;, \\
 F_{TEB}(\vec l_1 ; \vec l_2 ; \vec l_3 )
& = & -2 \Big\langle \frac{\delta T(\vec l_1)}{T} \;\kappa(\vec l_1)^* \Big\rangle^\prime
\sin(2 \varphi_{12}) C_{l_2}^{EE} \;.
\end{eqnarray}
The factor of 2 difference between the two expressions comes from the two possible $E$ to $B$ conversions of the $EE$ correlator. Note that 
here we do not have any derivative of the short wavelength spectrum: the effect of the long mode is to create a $B$ polarization and not to just
stretch or shrink a preexisting 2-point function.

The signal to noise ratios are   
\begin{equation}
\label{eq:SNB}
(S/N)_{TTB}^2 = \frac1\pi \int \frac{d^2 l_1 d^2 l_2}{(2 \pi)^2} \frac{F_{TTB}(\vec l_1 ; \vec l_2 ; \vec l_3)^2}{2 C^{TT}_{l_1}C^{TT}_{l_2}C^{BB}_{l_3}} \qquad 
(S/N)_{TEB}^2 = \frac1\pi \int \frac{d^2 l_1 d^2 l_2}{(2 \pi)^2} \frac{F_{TEB}(\vec l_1 ; \vec l_2 ; \vec l_3)^2}{C^{TT}_{l_1}C^{EE}_{l_2}C^{BB}_{l_3}} \;.
\end{equation}
For the variance $C^{BB}_l$ we take the one created by the lensing itself. The smallness of the $BB$ 2-point function gives a bigger signal to noise to these
correlators involving $B$ modes with respect to the others, at least in the case of an ideal experiment, limited only by cosmic variance. Moreover as the $BB$
2-point function coming from lensing peaks at $l \sim 1000$, the $S/N$ rises fast and then saturates around $l \sim 1500$.

\begin{figure}[!!ht]             
\begin{center}
\includegraphics[width=12cm]{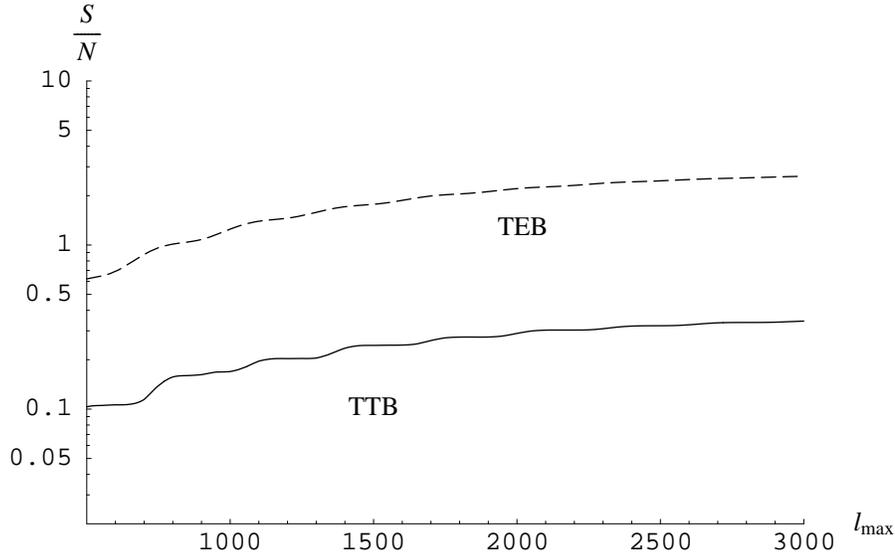}
\caption{\label{fig:SN4} \small Signal to noise for the TTB (continuous line) and TEB (dashed line) correlators as a function of the maximum $l$ for an ideal, 
cosmic variance limited experiment. Cosmological parameters are the same as in the previous figures.}
\end{center}
\end{figure}

\section{\label{sec:conc}Conclusions}
In this paper we calculated the 3-point function for the CMB temperature and polarization maps generated by non-linearities at recombination, 
in the limit in which one of the modes is outside the horizon at recombination and the other two are of much smaller length. These generated 3-point 
functions, besides their intrinsic interest, represent some sort of ``background'' if we are interested in studying the non-Gaussianity of the primordial 
spectrum of perturbations. Even though different models
predict different shape dependences for the primordial 3-point function \cite{Babich:2004gb}, usually experimental limits are set on the scalar variable 
$f_{\rm NL}$ defined through the relation
\begin{equation}
\label{eq:f_NL}
\zeta(x) =\zeta_g(x) -\frac35 f_{\rm NL}(\zeta_g(x)^2 - \left<\zeta_g^2\right>) \;,
\end{equation}  
where $\zeta$ is the observed perturbation and $\zeta_g$ is gaussian (\footnote{The non-linearity parameter is defined for the Newtonian 
potential in matter dominance; this explains the factor of 3/5 in the previous expression.}). The present limits on $f_{\rm NL}$ are given by the WMAP 
experiment $-58 < f_{\rm NL} < 134$ at $95\%$ of CL \cite{Komatsu:2003fd}. The Planck experiment will improve this limit down to $f_{\rm NL} \sim 5$. 

A useful way to express our result on the signal to noise of the TTT 3-point function is to calculate the equivalent $f_{\rm NL}$ which gives the same
signal to noise. Summing only over the triangles for which our approximation applies up to $l_{\rm max} = 3000$ we get 
\begin{equation}
\label{eq:FNLequiv}
f_{\rm NL}^{\rm equiv.} \sim 0.7 \;.
\end{equation}
This means that this non-Gaussianity should {\em not} be detected by Planck. Any signal of 3-point function (besides the well known ISW/lensing correlation) 
should be considered of primordial origin, at least in the geometrical limit in which our calculation applies. 
It is possible that future experiments will detect the 3-point functions generated by non-linearities at recombination especially using polarization modes, 
for which we obtained a bigger signal to noise ratio.

We stress that the configuration dependence of our result is completely different from that implied by eq.~(\ref{eq:f_NL}) which, in the geometrical limit 
we are considering, is proportional to the 2-point spectrum of the short modes. 
Our 3-point function, on the other hand, oscillates around zero as a function of the length of the short modes, being essentially proportional to the 
{\em derivative} of the 2-point spectrum. That is to say the non-Gaussian signature produced by the $f_{\rm NL}$ 
model in the presence of a long wavelength mode is a change in the amplitude of the 2-point function on small scales, 
a shift up and down of the $C_l$ curve. The effect we studied here changes the angular scales of the features in the 2-point function, a sideways shift in the 
$C_l$ curve.  Moreover, as the anisotropic effect is dominant, the 3-point function depends on the relative orientation between the long and short modes.
These differences should further reduce the contamination if we are looking for a primordial $f_{\rm NL}$.

Even though our calculation was done in a particular geometrical limit for the modes entering in the 3-point function, the signal to noise
we get summing over these particular squeezed configurations is not very small with respect to what is expected for the full 3-point function.
The reason is that, in the limit of a scale invariant 2 and 3-point function, the signal to noise of eq.~(\ref{eq:SN}) diverges by dimensional analysis
as $l_{\rm max}^2$ (in eq.~(\ref{eq:SN}) $F \propto l^{-4}$ and $C \propto l^{-2}$), with a possible additional log divergence which depends on the 
particular form of the 3-point function. Even restricting to the limit in which our results apply $S/N$ goes as $l_{\rm max}^2$ so that we expect to 
lose only a logarithmic factor by our inability to do the full calculation. This estimate does not consider factors of order one given by the physics at 
recombination which could enhance or suppress substantially $S/N$ in the regions in which our approximation is not valid.   

References \cite{Bartolo:2003gh, Bartolo:2003bz} calculate the TTT 3-point function in the limit in which all modes are outside the horizon at recombination. 
The result should be compatible with ours in the limit of squeezed triangles, with one side much bigger that the others, with 
all the modes outside the horizon. The 3-point function should go to zero because the spectrum is flat at large scales so that its derivative 
entering in eqs (\ref{eq:3-point1}), (\ref{eq:3-point2}) and (\ref{eq:3-point3}) vanishes. The result of ref.s \cite{Bartolo:2003gh, Bartolo:2003bz} does not 
show this behaviour. We were unable to trace back the origin of the disagreement between the results as the two approaches to the problem are 
completely different.

\section*{Acknowledgments}
We thank Wayne Hu, Juan Maldacena, Sabino Matarrese, Shinji Mukohyama, Antonio Riotto and Leonardo Senatore for useful discussions. 
M.~Z.~ is supported by NSF grants AST 0098606  and by the David and Lucille Packard Foundation Fellowship for Science and Engineering.

\appendix

\section*{Appendix A. Gauge invariance}
The 3-point function is obtained in the main text starting from the variation of the 2-point function moving along the background mode. The reader
could be worried about this procedure because it assumes that we can define the 2-point function before the long wavelength mode reenters in the horizon.
The usual derivation of the Sachs-Wolfe effect involves an integral over the photon trajectory from the last scattering surface to us. This seems
to indicate that the temperature fluctuation cannot be defined until the present time.

This is not quite true. Imagine that we have friends in different directions who look outwards along our line of sight. Imagine that they are 
sufficiently close to our last scattering surface to observe the 2-point function of the short modes on a small patch of it, just before the background mode 
reenters. We should specify which is the state of motion of our friends or, equivalently, the coordinate system with respect 
to which they are stationary. In the unperturbed Universe we could take them to be stationary with respect to the cosmic fluid, but including perturbations, 
different gauge choices will give a different relative velocity of order $\phi$ to our set of friends. This will only change at leading
order the dipole they observe, but if we are interested in higher multiples the result is unique and gauge independent. All this set of friends will
observe, in their different patch in the sky, a 2-point function with the {\em same} statistical properties because the background mode is still outside the horizon 
and therefore irrelevant: it just induces an unobservable rescaling of the spatial coordinates between one friend and another. From this point of view 
the effect of the background mode is to induce a deformation between the map one of our friends observed and the map of the {\em same} patch on the last scattering 
surface we observe now. The effect is physical and gauge invariant.

Thus the full effect is gauge invariant, but as we stressed in the main text, the separation of the various effects we discussed is not physical and 
depends on the gauge we choose. 
To show this in a concrete example let us start from the conformal Newtonian gauge (\ref{eq:Newton}) and redefine the time slicing to make it comoving, without
touching the space coordinates (total matter gauge). In matter dominance the time shift required is given by
\begin{equation}
\label{eq:timeshift}
\frac{\delta\tau}{\tau} = \frac\phi3 \;.
\end{equation}
The transformed metric is given as a function of the new time variable $\tilde\tau$ by
\begin{equation}
\label{eq:newm}
ds^2 = a^2(\tilde\tau) \left[-d\tilde\tau^2 + \Big(1-\frac{10}3\phi(x^i)\Big)dx_i dx_i +\frac23 \tilde\tau \nabla_i\phi \;d\tilde\tau d x_i\right] \;.
\end{equation} 
Outside the horizon, neglecting the gradient of $\phi$ and using the relation $\phi = -\frac35\zeta$, we get the metric (\ref{eq:comoving}) with
$\tilde\tau = \eta$.

As the gauge we are considering is comoving, there is no time redefinition effect in this metric; recombination happens at the same time $\tilde\tau_r$ 
everywhere. As there is no redefinition of spatial coordinates, the space redefinition and lensing effects are obviously left 
untouched. Therefore the time redefinition  effect must be incorporated in the Shapiro effect in this gauge. It is easy to verify this. At leading order in 
$\phi$ and neglecting lensing the null geodesics in the metric (\ref{eq:newm}) follow the equation
\begin{equation}
\label{eq:null}
d\tilde\tau = (1-\frac53\phi +\frac13 \tau\nabla\phi) dx \;.
\end{equation}
This gives integrating by parts
\begin{equation}
\tilde\tau - \tilde\tau_r = \int (1-2\phi(x')) dx' - \tilde\tau_r \frac\phi3 (x)\;,
\end{equation}
which coincides with equation (\ref{eq:geodesics}). The fluctuation along the background wave of the time of recombination in a generic gauge becomes 
a contribution to the distance to the last scattering surface in a comoving gauge.

The lensing effect is, on the other hand, independent of the parameterization of the perturbations. Propagating backwards from us the null geodesics, 
we can covariantly define the shear tensor on a surface perpendicular to the line of sight as the pull back on the surface of the covariant derivative \cite{Wald:rg}
\begin{equation}
\label{eq:nullshear}
\nabla_\mu k_\nu \;,
\end{equation}
where $k_\mu$ is the null vector field tangent to the geodesics. As we are interested in the result at first order in the perturbation, both the
surface perpendicular to the line of sight and the affine parameterization of the geodesics can be made in the unperturbed Universe and it is
therefore common to all gauges. The integral of the shear tensor along the line of sight is therefore gauge independent.

\section*{Appendix B. Expressions on the full sky}
In the main text we discussed our results in the flat sky limit, which is much more transparent. In this appendix we generalize some of the expressions to 
full sky. The correspondence between flat sky and full sky for the 3-point functions is thoroughly studied in \cite{Hu:ee}.

In a full sky analysis definition (\ref{eq:defF}), which in the flat sky limit enforces the translational invariance of the 3-point function, is replaced by 
\begin{equation}
\label{eq:wigner}
\langle (\delta T/T)_{l_1 m_1} (\delta T/T)_{l_2 m_2} (\delta T/T)_{l_3 m_3} 
\rangle = \left(\begin{array}{ccc} l_1 & l_2 & l_3 \\ m_1 & m_2 & m_3 \end{array}\right)  B(l_1,l_2,l_3) \;, 
\end{equation}
involving the Wigner $3j$ symbols, which is manifestly rotationally invariant.

It is straightforward to extend the flat sky 3-point functions to the full sky. In the case of isotropic contributions, in which there in no
dependence on the relative angle between the long and the short modes
\begin{equation}
\label{eq:isofull}
B_{\rm iso}(l_1,l_2,l_3) = \left(\begin{array}{ccc} l_1 & l_2 & l_3 \\ 0 & 0 & 0 \end{array}\right) \sqrt{\frac{(2l_1+1)(2l_2+1)(2l_3+1)}{4\pi}} 
F(\vec l_1 ; \vec l_2 ; \vec l_3 ) \;,
\end{equation}
where $F$ in eq.s (\ref{eq:3-point1}) and (\ref{eq:3-point2}) must be written as product of the two full sky spectra. The previous expression
is different from zero only for $l_1+l_2+l_3$ even, as implied by parity invariance.

For the anisotropic contribution eq.~(\ref{eq:3-point3}) the cosine of the angle between the long and short modes goes into the Wigner $3j$
symbols giving
\begin{equation}
\label{eq:anyfull}
B_{\rm aniso}(l_1,l_2,l_3) = \left(\begin{array}{ccc} l_1 & l_2 & l_3 \\ 2 & 0 & -2 \end{array}\right) \sqrt{\frac{(2l_1+1)(2l_2+1)(2l_3+1)}{4\pi}} 
\Big\langle \frac{\delta T}{T} \;\kappa^* \Big\rangle_{l_1}
\frac1{l_2^2} \left(-\frac{\partial}{\partial \log l} (l_2^2 C_{l_2})\right)\;.
\end{equation}
This expression holds only for $l_1+l_2+l_3$ even, while it is taken to be zero for parity violating correlations.
The same correspondence holds for the polarization spectra $TTE$ and $TEE$ eq.~(\ref{eq:3-pointE}). For the 3-point functions involving $B$ modes
we have 
\begin{equation}
\label{eq:Bfull}
B_{TTB}(l_1,l_2,l_3)
= i \left(\begin{array}{ccc} l_1 & l_2 & l_3 \\ 2 & 0 & -2 \end{array}\right) \sqrt{\frac{(2l_1+1)(2l_2+1)(2l_3+1)}{4\pi}}
\Big\langle \frac{\delta T}{T} \;\kappa^* \Big\rangle_{l_1} C_{l_2}^{TE} \;
\end{equation}
and similarly for the $TEB$ correlator. In this case the previous expression holds for $l_1+l_2+l_3$ odd, while it is taken to be zero in the other cases.

In the full sky the integral over the line of sight which gives the temperature-convergence correlation (see eq.~(\ref{eq:kappaphi}) for the flat sky expression) 
takes the form
\begin{equation}
\label{eq:fullkappaT2}
\Big\langle (\delta T/T)_{l_1 m_1} \;\kappa^*_{\;l_2 m_2} \Big\rangle = \delta_{m_1 m_2}\delta_{l_1 l_2} \cdot \frac3{25} (-4 \pi) l(l+1) P_{\cal{R}} \int_0^\infty 
\frac{dy}y j_{l_1}(y) \int_0^1 \frac{dx}x (1-x) j_{l_1}(x y) \;,
\end{equation}
where $j$ are spherical Bessel functions. In the spherical geometry we do not have the problem we discussed in the main text about the correspondence 
between the transverse momentum and the mode we observe in the sky. We can check that the flat sky approximation used in the main text is quite good, except
for the lowest multiples.

\end{document}